\documentclass[prl, twocolumn,superscriptaddress,showpacs,amsmath,amssymb,14pt]{revtex4}

\usepackage{graphicx}
\usepackage{dcolumn}
\usepackage{bm}
\usepackage{color}
\usepackage{hyperref}
\usepackage{multirow}
\newcommand{\sitn}{$^{29}$Si}
\newcommand{\ket}[1]{\left| #1 \right\rangle}
 
\begin{document}

\title{ Coherent storage of photoexcited triplet states using  
$^{29}$Si nuclear spins in silicon}

\author{Waseem Akhtar}
\affiliation{ School of Fundamental Science and Technology, Keio University, 3-14-1 Hiyoshi, Kohuku-ku, Yokohama 223-8522,Japan }

\author{Vasileia Filidou}
\affiliation{Department of Materials, University of Oxford, Parks Road, Oxford OXI 3PH, UK }

\author{Takeharu Sekiguchi}
\affiliation{ School of Fundamental Science and Technology, Keio University, 3-14-1 Hiyoshi, Kohuku-ku, Yokohama 223-8522,Japan }

\author{Erika Kawakami}
\affiliation{ School of Fundamental Science and Technology, Keio University, 3-14-1 Hiyoshi, Kohuku-ku, Yokohama 223-8522,Japan }

\author{Tatsumasa Itahashi}
\affiliation{ School of Fundamental Science and Technology, Keio University, 3-14-1 Hiyoshi, Kohuku-ku, Yokohama 223-8522,Japan }

\author{Leonid Vlasenko}
\affiliation{ A. F. Ioffe Physico-Technical Institute, Russian Academy of Sciences,194021, St. Petersburg,  Russia}

\author{John J. L. Morton}
\email{john.morton@materials.ox.ac.uk}
\affiliation{Department of Materials, University of Oxford, Parks Road, Oxford OXI 3PH, UK }
\affiliation{CAESR, The Clarendon Laboratory, University of Oxford, Parks Road, Oxford OXI 3PU, UK}

\author{Kohei M. Itoh}
\email{kitoh@appi.keio.ac.jp}
\affiliation{ School of Fundamental Science and Technology, Keio University, 3-14-1 Hiyoshi, Kohuku-ku, Yokohama 223-8522,Japan }



\date{\today}

\begin{abstract}
Pulsed electron paramagnetic resonance spectroscopy of the photoexcited, metastable triplet state of the oxygen-vacancy center in silicon reveals that the lifetime of the $m_s$=$\pm$1 sub-levels differ significantly from that of the $m_s$=0 state. We exploit this significant difference in decay rates to the ground singlet state to achieve nearly $\sim$100$\%$ electron spin polarization within the triplet. We further demonstrate the transfer of a coherent state of the triplet electron spin to, and from, a hyperfine-coupled, nearest-neighbor \sitn~nuclear spin. We measure the coherence time of the $^{29}$Si nuclear spin employed in this operation and find it to be unaffected by the presence of the triplet electron spin and equal to the bulk value measured by nuclear magnetic resonance.     
\end{abstract}

\pacs{76.30.-v, 71.55.Cn, 76.70.Dx, 03.67.Lx}

\maketitle



Nuclear spins in solids are promising candidates for quantum bits (qubits) as their weak coupling to the environment often leads to very long spin coherence times~\cite{LGY,KM, LMY,MTB}.
However, performing fast manipulation and controlling interaction between nuclear spin qubits is often more challenging than in other, more engineered, quantum systems~\cite{PJT,NSL,OPT}.
The use of an optically driven mediator spin has been suggested as a way to control coupling between donor electron spins in silicon: the donor spins exhibit weak direct coupling, but mutually couple through the optically excited state of the mediator~\cite{SFG}. Such ideas could similarly be applied to couple nuclear spins, and, if the mediator spin is a photo-excited triplet with a spin-zero single ground state, it would have the added advantage that it avoids long-term impact on the nuclear spin coherence~\cite{SFK, GRS, SLG}.

Photoexcited triplets are optically-generated electron spins ($S=1$) which often exhibit large (positive or negative) spin polarization, thanks to preferential population of each of the triplet sub-levels following intersystem crossing  and/or the differing decay rates of these sub-levels to the ground singlet state~\cite{JH,DBD}. Nuclear spins, in contrast, have weak thermal spin polarization at experimentally accessible conditions, due to its small magnetic moment. Highly polarized electron spin triplets can be used to polarize surrounding nuclear spins, through continuous wave microwave illumination (under processes termed dynamic nuclear polarization)~\cite{BBH,TTT}, or using microwave pulses~\cite{SHH}. Triplet states can also be used to mediate entanglement between mutually-coupled nuclear spins~\cite{SFK}, on timescales much faster than their intrinsic dipolar coupling~\cite{VSinprep}.


Oxygen-vacancy (O-$V$) complexes can be formed in silicon by electron beam or $\gamma$-ray  irradiation of oxygen-rich silicon crystals~\cite{WCW,CWC}, and can be excited to the triplet state (termed an SL1 center,) using illumination of above band gap light~\cite{KB}. 
Magnetic resonance studies including electron paramagnetic resonance (EPR), electrically or optically detected magnetic resonance, spin dependent recombination and electrically-detected cross relaxation~\cite{KB,SGS,WMV,WM,VMG, AMS} have revealed that the SL1 has orthorhombic symmetry with non-equilibrium triplet electron spin polarization and strong hyperfine coupling with the nearest neighbor $^{29}$Si nuclear spins. 
The triplet spin polarisation of SL1 centers can be incoherently transferred to bulk $^{29}$Si nuclear spins in the lattice by all-optical methods~\cite{LV} or dynamic nuclear polarization~\cite{VVP,IHI}. 
Previous electron spin echo studies on SL1 performed at zero magnetic field have revealed that the populating rates of the triplet sub-levels are equal, and the spin polarization arises instead from a difference in the decay rates to the singlet ground state~\cite{FBvO}.
In this Letter we use the high spin polarization of the triplet system and its strong coupling with the nearest neighbor $^{29}$Si nuclear spins to demonstrate coherent state transfer between the electron and nuclear spin degrees of freedom, and examine the nuclear spin coherence in the presence of the triplet.
  
Cz-grown, single-crystal natural silicon (4.7\% $^{29}$Si, $I=1/2$) was exposed to 1~MeV e-beam irradiation (dose $\approx$ 10$^{18}$ cm$^{-2}$) at room-temperature to form O-$V$ complexes (an interstitial oxygen already present in the silicon traps a mono-vacancy generated due to the e-beam irradiation). Pulsed EPR measurements were carried out at X-band (9.72~GHz) on a Bruker Elexsys580 spectrometer equipped with a helium-flow cryostat. 
Photoexcitation of the SL1 was achieved using a 1064~nm pulsed Nd:YAG laser (pulse width $\sim 7$~ns, 1~mJ/pulse) with a 10 Hz repetition rate.

\begin{figure}[t]
\begin{center}
\includegraphics[width=8.5cm]{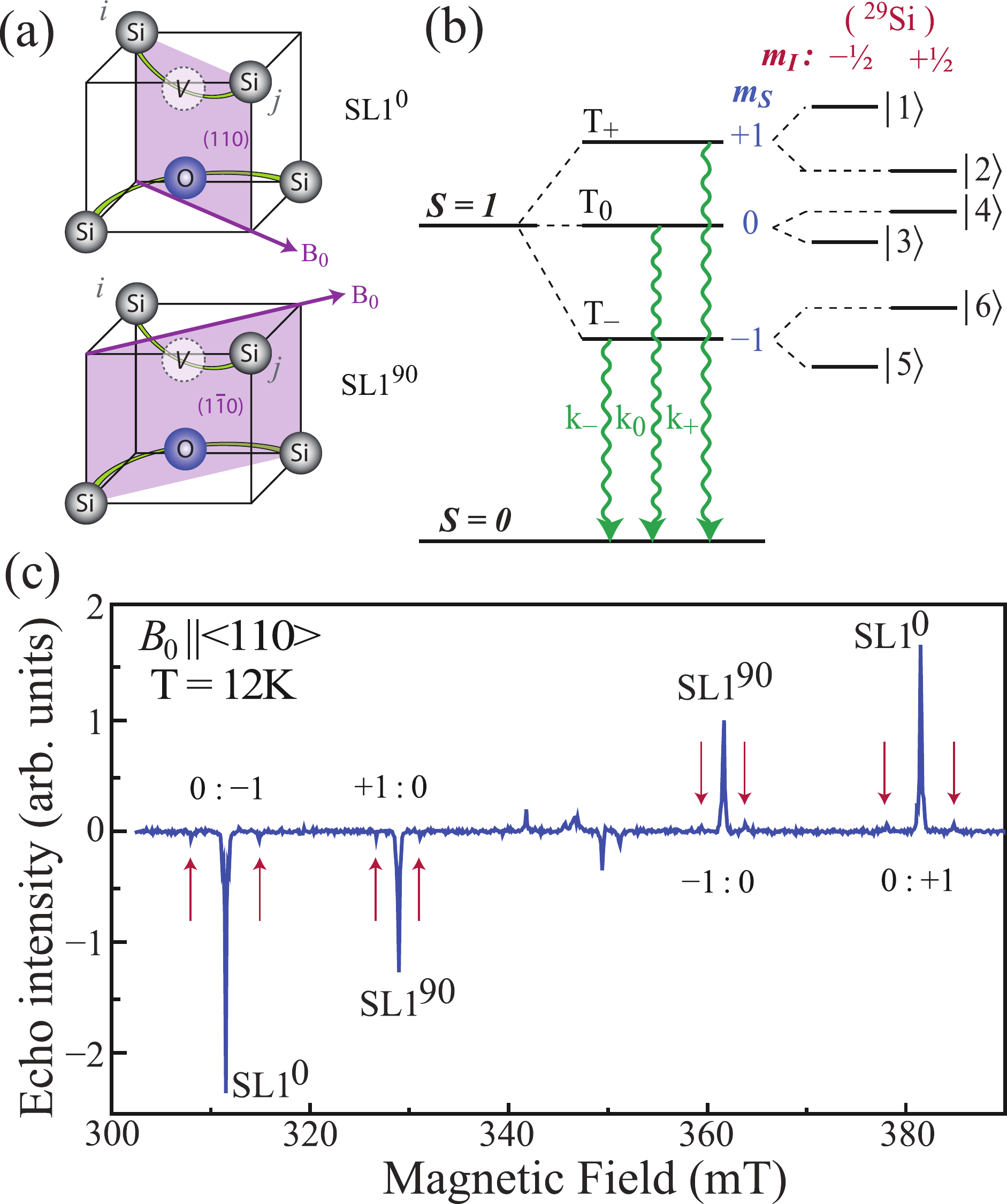}
\caption {(color online). (a) Structures of the oxygen-vacancy centre in silicon, illustrating the SL1$^0$ and SL1$^{90}$ orientations with respect to the externally applied magnetic field $B_0$. (b) The SL1 triplet ($S=1$) state is Zeeman split by $B_0$ into levels T$_+$, T$_0$, and T$_-$. These states decay with different rates ($k_+$, $k_0$, and $k_-$, respectively) to the ground singlet ($S=0$) state. The hyperfine coupling to the  $^{29}$Si ($I=1/2$) nuclear spin at lattice site  $i$ or $j$ further splits the triplet states. (c) Electron spin echo-detected EPR spectrum of the SL1 center at 12~K with $B_0 \| \left<110\right>$. The satellite peaks (red arrows) arise from hyperfine coupling to $^{29}$Si.}  
\label{fig:one}
\end{center}
\end{figure}

Figure~\ref{fig:one}a illustrates the SL1 center in silicon under two representative orientations of the static magnetic field $B_0$. Under one orientation, termed SL1$^0$, the magnetic field lies in the plane, marked (110), comprising the oxygen atom and two vacancy-trapping silicon atoms ($i$ and $j$ lattice sites). An alternative orientation (SL1$^{90}$) has the magnetic field in an orthogonal plane (1$\overline{1}$0) with respect to the same center. Both planes are equivalent by symmetry for the crystal as a whole, so both orientations are visible in the EPR spectrum.

Figure~\ref{fig:one}b shows the triplet energy sub-levels in the presence of a static magnetic field (T$_+$, T$_0$, and T$_-$), each of which decays with a characteristic rate 
to the ground singlet state~\cite{VMG,VZK}. 
The hyperfine coupling with one of the two nearest-neighbor $^{29}$Si nuclear spins (occupying $i$ or $j$ lattice site) further split the T$_{\pm1}$ sub-levels, while the T$_0$ state   ($m_s=0$) has no first-order hyperfine interaction, thus the nuclear spin splitting in this sub-level is close to the Zeeman energy of $^{29}$Si. The EPR spectrum obtained by monitoring the electron spin echo intensity as a function of magnetic field at 12 K with $B_0 \| \left<110\right>$ is shown in Fig.~\ref{fig:one}c, labeled with electronic transitions identified by previous cw-EPR studies~\cite{KB}. The satellite peaks accompanying each main peak are due to the hyperfine interaction with the $^{29}$Si nuclear spins situated at $i$ or $j$ lattice sites. 
The phase difference of the spin echo (i.e.\ dips or peaks) is indicative of the non-equilibrium polarization within the electron spin triplet.

\begin{figure}[t]
 \begin{center}
 \includegraphics[width=8.5cm]{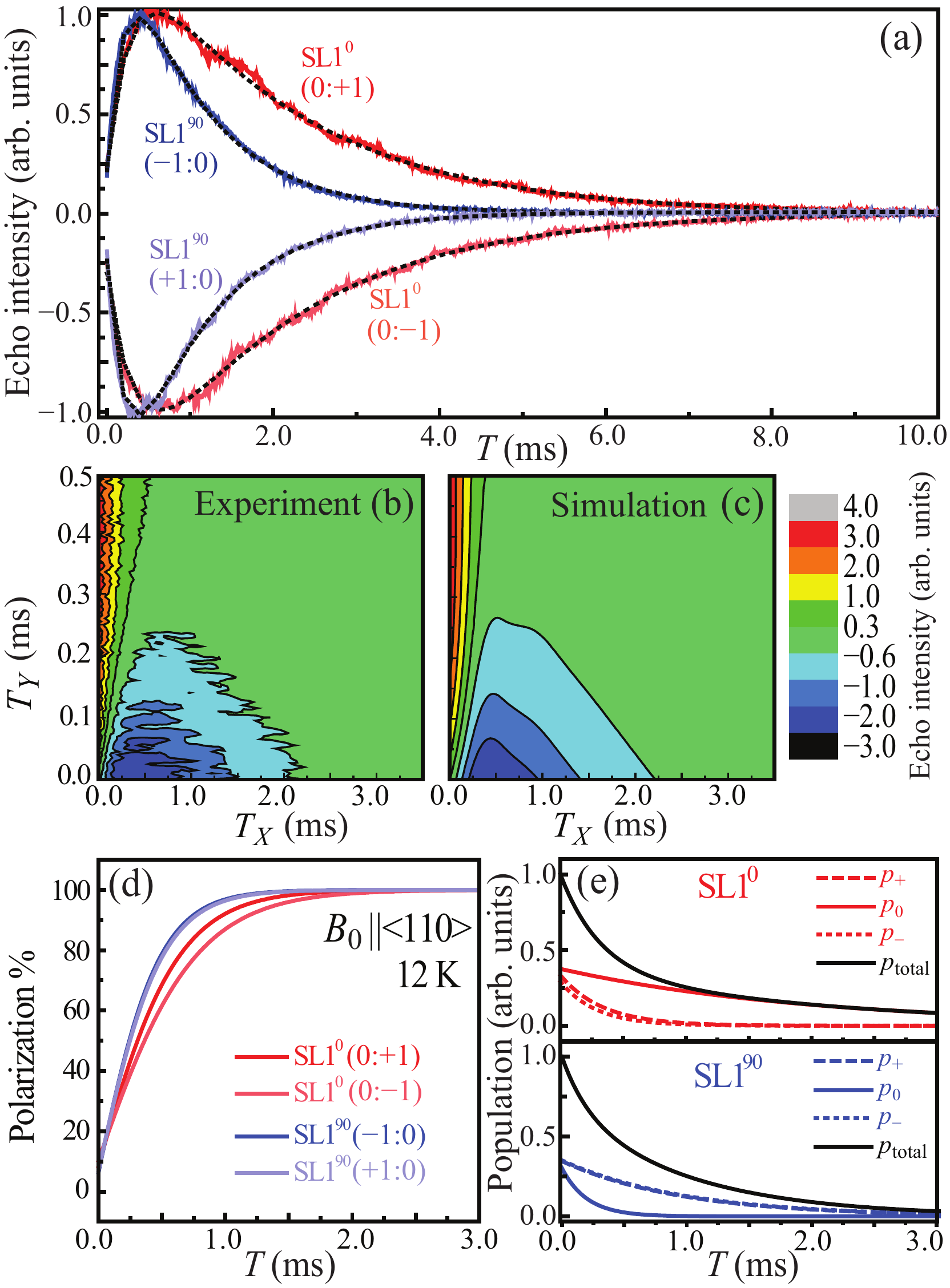}
 \vspace*{-.2cm}
 \caption {(color online). (a) Decay traces obtained by the flash delay ($h\nu$-$T$-$\pi/2$-$\tau$-$\pi$-$\tau$-$echo$) pulse sequence, used to extract the triplet decay rates shown in Table~\ref{table:nonlin}. (b) Experimental and (c) simulated 2D plots of the decay characteristics observed with the  pulse sequence $h\nu$-$T_X$-$\pi$-$T_Y$-$\pi/2$-$\tau$-$\pi$-$\tau$-$echo$. (d) The polarization build-up and (e) triplet population as a function of waiting time $T$ after the laser pulse.} 
 \label{fig:two}
 \end{center}
 \end{figure}

To investigate the origin of this non-equilibrium polarization, we studied the decay kinetics of the triplet by measuring the electron spin echo at a variable time $T$ after the optical excitation ($h\nu$-$T$-$\pi/2$-$\tau$-$\pi$-$\tau$-$echo$), as shown in Fig.~\ref{fig:two}a. 
The zero echo intensity at $T = 0$ 
indicates equal initial filling of the three triplet sub-levels upon creation of the triplet.
The echo intensity proceeds to grow as $T$ is increased. This can be attributed to a difference in the decay rates of the triplet sub-levels to the ground state, creating a build-up in spin polarisation (positive or negative) across the EPR transition being measured. 

Based on the simple decay model shown in Fig.~\ref{fig:one}b, the population difference between a given pair of sub-levels follows a biexponential behavior where the two time constants represent the lifetimes of the two sub-levels involved in the EPR transition. The time constants obtained from biexponential fitting to the FD curves are given in Table~\ref{table:nonlin} --- the assignment of rates to particular energy levels is enabled by electron nuclear double resonance experiments described further below. As the transition from the triplet to the ground singlet is determined by the amount of singlet admixture to the triplet via spin-orbit coupling, the lifetimes are expected to depend on the defect orientation with respect to the magnetic field. The composition of tripet levels T$_{+,0,-}$ can be expressed in terms of the zero-field eigenstates (T$_{x,y,z}$), and similarly the observed decay rates from these levels can be traced back to a corresponding mixture of zero-field decay rates (k$_{x,y,z}$), as shown in Table 1. These values are in good agreement with times measured using zero-field EPR~\cite{FBvO}.

Our model assumes that there is negligible spin-lattice relaxation within the triplet sub-levels, and this is consistent with lack of temperature dependence we observe in the relaxation dynamics below 20~K. Neverthless, in order to probe the dynamics in more detail, we can introduce an additional inversion $\pi$ pulse to the sequence: ($h\nu$-$T_X$-$\pi$-$T_Y$-$\pi/2$-$\tau$-$\pi$-$\tau$-$echo$). Figure~\ref{fig:two}c shows the 2D plot of the echo intensity for this sequence, as both $T_X$ (the delay after the laser pulse) and $T_Y$ (the delay after the inversion pulse) are varied. The simulation of this experiment, based on the model described above, is in good agreement with the observed behavior, supporting our assumption that spin-lattice relaxation can be neglected.

Based on the observed decay rates, we can extract both the polarization buildup (Fig.~\ref{fig:two}d) and the triplet population (Fig.~\ref{fig:two}e) for the two SL1 orientations (SL1$^0$ and SL1$^{90}$) as a function of time $T$ after the laser pulse. The maximum electron polarization reaches $>99\%$ after about 1.5 ms following the laser pulse. Below we investigate the coherent transfer of such well-prepared electron spin states to a neighboring $^{29}$Si nuclear spin.

\begin{table}[ht]
\caption{Lifetime of triplet sub-levels, for two orientations of SL1 centers, obtained from fitting to the flash delay curve.} 
\centering 
\begin{tabular}{c|c } 
\hline 
\hline
EPR transition  & Lifetime \\ 
\hline
$0\rightarrow+1$, SL1$^0$ & {(1/k$_0$)$^0$}= 2000(4) $\mu$s, {(1/k$_{+1}$)$^0$}= 280(2) $\mu$s \\ [1ex]   
$0\rightarrow-1$, SL1$^0$ & {(1/k$_0$)$^0$}= 1970(4) $\mu$s, {(1/k$_{-1}$)$^0$}= 330(2) $\mu$s \\ [1ex] 
\hline
$-1\rightarrow0$, SL1$^{90}$ & {(1/k$_{-1}$)$^{90}$}= 960(2) $\mu$s, {(1/k$_{0}$)$^{90}$}= 200(1) $\mu$s \\ [1ex]   
$+1\rightarrow0$, SL1$^{90}$ & {(1/k$_{+1}$)$^{90}$}= 987(2) $\mu$s, {(1/k$_{0}$)$^{90}$}= 205(1) $\mu$s \\ 
\hline
\hline
\multicolumn{2}{c}{$k_x=1.6(3)$~ms$^{-1}$, $k_y=4.93(6)$~ms$^{-1}$, $k_z=0.50(4)$~ms$^{-1}$}  \\
\hline
\hline
\end{tabular}
\label{table:nonlin} 
\end{table}

\begin{figure}[t]
\begin{center}
\includegraphics[width=8.7cm]{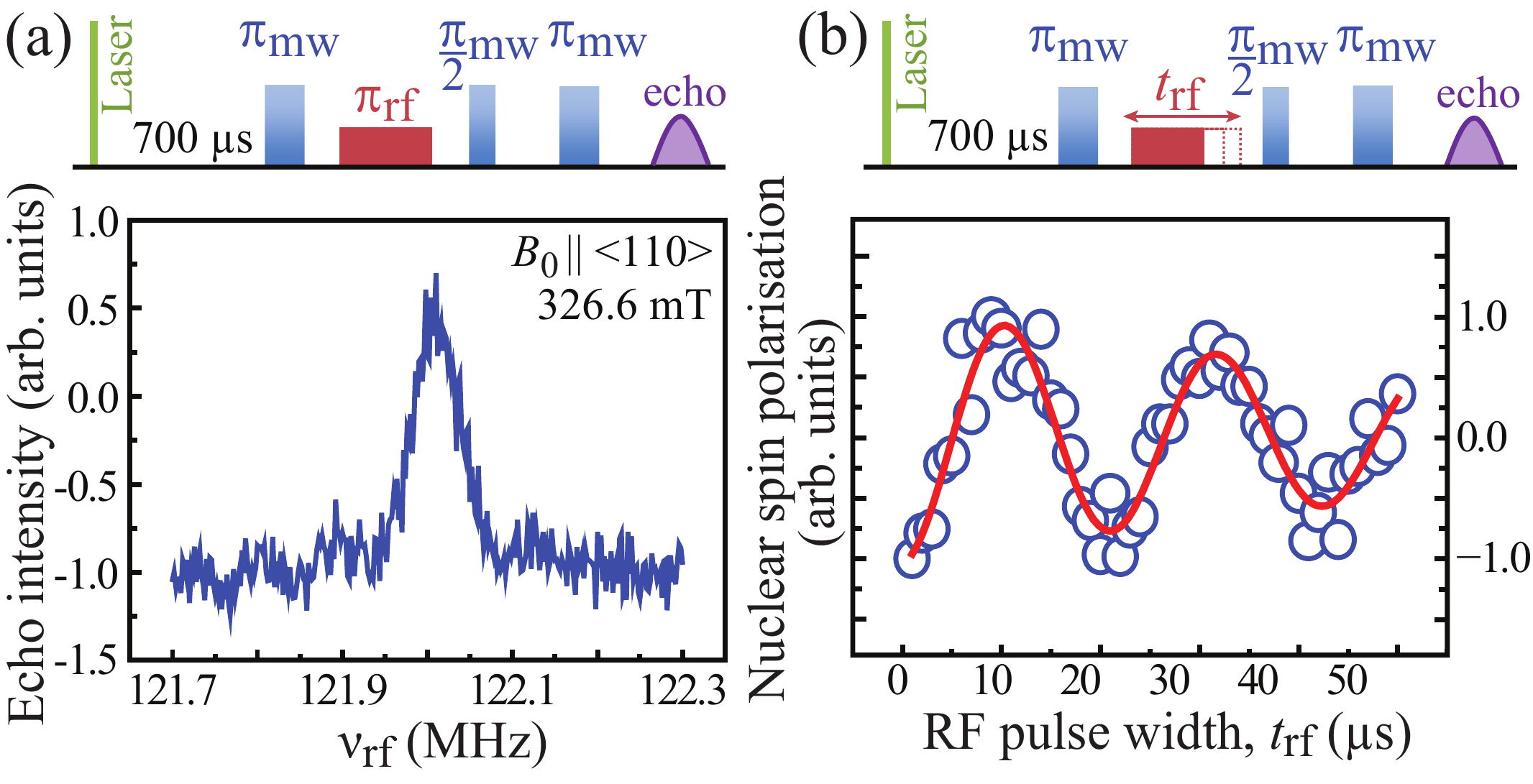}
\vspace*{-.2cm}
\caption {(color online).  (a) Davies-ENDOR spectrum illustrating the hyperfine coupling between the triplet and nearest-neighbour \sitn~for the SL1$^{90}$ center. (b) Rabi oscillation of the $^{29}$Si nuclear spin, driven between the states $\ket{1}$ and $\ket{2}$, detected by monitoring the electron spin echo intensity.}
\label{fig:three}  
\end{center}
\end{figure}

For the following electron nuclear double resonance (ENDOR) experiments, we focus on four selected levels (labeled $\ket{1},\ket{2},\ket{3}$ and $\ket{4}$ in Fig.~\ref{fig:one}b) of SL1$^{90}$. 
We studied the hyperfine coupling strength between the $^{29}$Si nuclear spin and triplet electron spin using the Davies ENDOR pulse sequence (Fig.~\ref{fig:three}a). 
For SL1$^{90}$, T$_0$ decays more quickly than T$_{\pm}$ (see Table~\ref{table:nonlin}), thus the states $\ket{3}$ and $\ket{4}$ are mostly unpopulated in $\sim$700 $\mu$s. 
A selective microwave $\pi$-pulse between $\ket{1}$ and $\ket{3}$ creates a polarization across the nuclear spin transitions, which can be driven using a radiofrequency ($\nu_\text{rf}$) pulse. The ENDOR signal is obtained by monitoring the electron spin echo on the $\ket{1}$:$\ket{3}$ transition as a function of $\nu_\text{rf}$.  Fig.~\ref{fig:three}a shows the $\ket{1}$:$\ket{2}$ transition frequency dominated by the strong hyperfine interaction.
Thus, with resonant rf pulses we can selectively address \sitn~nuclear spins at specific lattice sites $i$ and $j$. These nuclear spins can be coherently manipulated with high fidelity, as illustrated by the Rabi oscillations in Fig.~\ref{fig:three}b, in addition to being prepared and measured using the triplet electron spin.

\begin{figure}[t]
 \begin{center}
 \includegraphics[width=8.5cm]{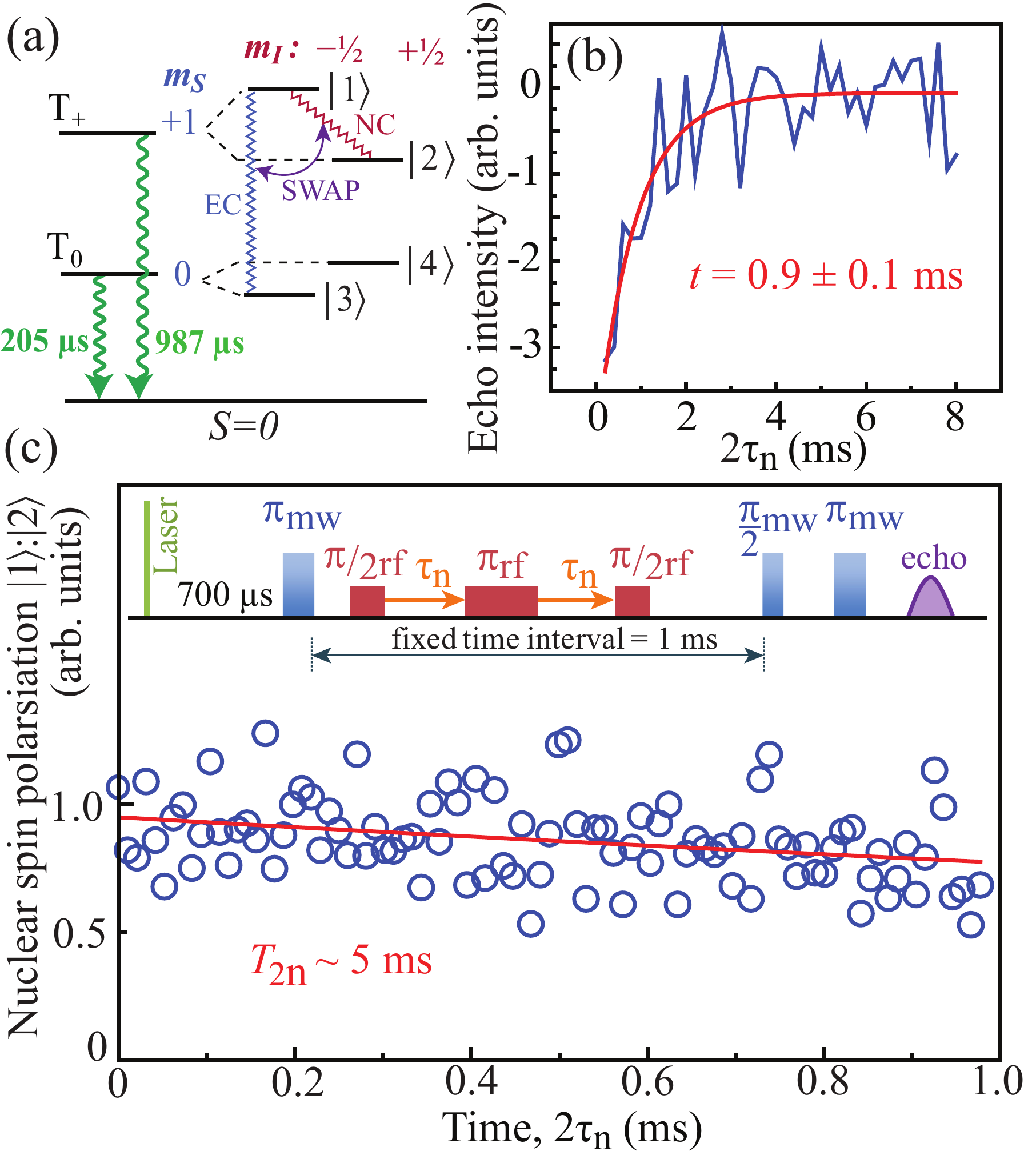}
 \vspace*{-.2cm}
 \caption {(color online). 
(a) The electron-spin coherence (EC) of the triplet is transferred to, and from, a \sitn~nuclear coherence (NC) . (b) The retrieved electron spin echo intensity (blue) decays as a function of the $^{29}$Si nuclear storage time, 2$\tau _n$. A fit to an exponential decay (red) yields a time constant of 0.9(1)~ms, apparently limited by the lifetime (0.99~ms) of the T$_+$ state. (c) By fixing the experimental time window to $\sim1$~ms, the intrinsic coherence time of the nuclear spin can be measured.
rf pulses were phase-cycled to confirm here that the measured electron spin echos arose solely from nuclear spin coherences.} 
\label{fig:four}
 \end{center}
 \end{figure}

Using a sequence based on Davies ENDOR, it is possible to coherently transfer a state of the electron spin to a coupled nuclear spin (see Ref~\cite{MTB} for full details). We apply this to transfer the highly polarised triplet electron spin coherence to, and from, the nearest neighbour \sitn~nuclear spin (see Figure~\ref{fig:four}a).
The decay of the recovered spin coherence as a function of the storage time in the nuclear spin (2$\tau _n$) is shown in  Fig.~\ref{fig:four}b, with an exponential decay of time-constant 0.9(1)~ms. 
The measured decay is dominated by the relaxation of the T$_+$ sub-level back to the ground singlet state ($1/{k_+}=987~\mu$s), rather than $^{29}$Si nuclear decoherence. This confirms our assignment of the decay rates shown in Table~\ref{table:nonlin}.
By subtracting $k_+$ from the decay in  Fig.~\ref{fig:four}b we can estimate $T_2$ of \sitn~to be several milliseconds, however, it is possible to make a more accurate measurement as described below.

Using the pulse sequence shown in Fig.~\ref{fig:four}c, we shift the RF pulses within a fixed experimental window of 1~ms in order to remove the effect of the triplet relaxation and directly measure the \sitn~coherence time~\cite{Hoefer}. 
The sequence is based on Davies ENDOR described above, but with the single rf $\pi$ pulse replaced with a nuclear Hahn echo sequence, whose delay time $\tau_n$ is swept. In the absence of nuclear spin decoherence, the applied rf pulses form a net 2$\pi$ rotation. In contrast, when the nuclear spin is fully decohered, the nuclear spin polarisation across the $\ket{1}$:$\ket{2}$ transition falls to zero.
Fitting the data to an exponential decay gives the nuclear coherence time of 5(1) ms.  The bulk value for $T_{\rm 2n}$ of $^{29}$Si in natural silicon has been measured by NMR and found to be 5.6~ms, limited by \sitn~dipolar coupling~\cite{Barrett}. Remarkably, the nuclear spin coherence appears unaffected by strong coupling to the triplet electron spin.

In conclusion, we utilized the coupling between nuclear spin and photoexcited electron spin triplet in silicon to demonstrate the coherent storage and retrieval of triplet electron spin coherence in the \sitn~nuclear spin. 
This motivates further studies in how the nuclear spin state survives the decay to the ground singlet state, as well as the application of NMR pulse sequences to remove the effect of nuclear spin dipolar couplings to achieve coherence times of up to 25 seconds~\cite{LMY}. Furthermore, given the well-developed silicon isotope engineering~\cite{TIO,IKU}
it will be possible to investigate more than one nuclear spin strongly coupled to the single electron spin in the photoexcited triplet state and explore optical control of the interaction between the nuclear spins.
  
This work was supported in part by Grant-in-Aid for Scientific Research and Project for Developing Innovation Systems by MEXT, FIRST, Keio G-COE and JST-EPSRC/SIC (EP/H025952/1), at Oxford by the EPSRC through CAESR (EP/D048559/1). J.J.L.M is supported by the Royal Society and St. John's College, Oxford.


\end{document}